\documentclass[%
 reprint,
nobibnotes,
 amsmath,amssymb,
 aps,
 prc,
]{revtex4-2}
\usepackage{graphicx}% Include figure files
\usepackage{dcolumn}% Align table columns on decimal point
\usepackage{bm}% bold math
\usepackage[hidelinks]{hyperref}% add hypertext capabilities
\usepackage{caption}
\usepackage{subcaption}
\usepackage{xcolor}
\usepackage{cancel}

\usepackage{color}

\begin{document}

\title{Decay-Resolved Charge Changes from Radioactive Decays in Levitated Microparticles}

\author{Jiaxiang Wang}
\altaffiliation{Present address: Department of Physics and Astronomy,
University of California, Los Angeles, 90095 Los Angeles, CA}
\author{T. W. Penny}
\altaffiliation{Present address: Department of Microtechnology and Nanoscience (MC2),
Chalmers University of Technology, 41296 Gothenburg, Sweden}
\author{Yu-Han Tseng}
\author{Benjamin Siegel} 
\author{David C. Moore}
\affiliation{%
Wright Laboratory, Department of Physics, Yale University, New Haven, Connecticut 06520, USA
}%

\date{\today}

\begin{abstract}
We measure event-by-event discrete changes in the net electric charge of an optically levitated silica microsphere arising from individual radioactive decays within the sphere, in coincidence with energy depositions in a nearby scintillation detector. The net charge of the levitated sphere is continuously monitored by measuring its driven response to an oscillating electric field, allowing individual charge-change events to be resolved on millisecond timescales with precision below an elementary charge. Simultaneously, $\alpha$ and $\beta$ particles emitted during decays of implanted $^{212}$Pb and its daughters are detected using a scintillator read out with an array of silicon photomultipliers. By correlating reconstructed charge-change times with the scintillator response, we can directly attribute abrupt changes in the sphere’s net charge to individual nuclear decays, and identify differences in the distribution of charges ejected for $\alpha$ and $\beta$ decays. These results establish a new approach for studying low energy charged particles emitted by radioactive decays at the single-decay level, and identify showers of radiogenically produced low-energy electrons emitted by $\alpha$-decaying radon daughters implanted near solid surfaces. 

\end{abstract}

\maketitle
\section{\label{sec:intro}Introduction}

Levitated optomechanical systems offer a sensitive platform for detecting small forces~\cite{carlos_levita_2021}. A micro- or nano-particle levitated in vacuum can exhibit exceptionally low mechanical dissipation and high isolation from environmental noise~\cite{gieseler_subkelvin_2012,li_millikelvin_2011, Martynas_impulse_2026}.
Due to the isolation from thermal noise and small masses possible with levitated systems, force sensitivities can reach the zeptonewton to yoctonewton scale~\cite{magrini_real-time_2021,Ranjit_Zeptonewton_2016,Hebestreit_sensing_2018,Liang_yoctonewton_2023}. 

Due to this sensitivity, levitated optomechanical sensors can provide a uniquely powerful platform for the detection of weak forces in a number of areas of fundamental physics. They have been proposed as probes for Casimir and non-contact frictional forces~\cite{manjavacas_vacuum_2010,zhao_rotational_2012} and high-frequency gravitational waves~\cite{aggarwal_searching_2022}. They have also been applied to search for weak, long-range interactions beyond the Standard Model between masses~\cite{venugopalan_optomechanical_2026} or from dark matter~\cite{Moore:2020QST_review,monteiro_search_2020,afek_coherent_2022,tseng_search_2025,hamaide_searching_2025}, and to search for millicharged particles~\cite{afek_limits_2021}. With further improvements in motional control~\cite{magrini_real-time_2021,tebbenjohanns_quantum_2021,Ranfagni_quantum_2022,piotrowski_simultaneous_2023,dania_high-purity_2025,troyer_quantum_2025}, optimizing measurement imprecision and back-action~\cite{maurer_quantum_2023,tebbenjohanns_optimal_2019,schreppler_optically_2014}, and quantum state manipulation~\cite{Uros_cooling_2020,militaru_ponderomotive_2022,Mitsuyoshi_quantum_2025}, such systems are poised to detect even smaller forces~\cite{Martynas_impulse_2026,lee_impulse_2025,Giacomo_squeeze_2026}, including potentially providing a platform for studying the behavior of gravitational interactions between quantum systems~\cite{Marshman_locality_2020,Scala_matter_2013,rossi_quantum_2025,bose_massive_2025}. Extending these systems to large arrays of particles can enable searches for even rarer interactions~\cite{afek_coherent_2022,Moore:2020QST_review} or enhanced sensing through collective effects in coupled arrays~\cite{brady_entanglement-enhanced_2023,siegel_optical_2025,reisenbauer_non-hermitian_2024,yokomizo_non-hermitian_2023,Rudolph_hermitian_2024,Rudolph_force_2022,Rudolph_quantum_2024}. 

Future applications of levitated sensors include reconstructing the full kinematics of the decay products of nuclear $\beta$ or electron capture decays, allowing the properties of the emitted neutrinos to be inferred~\cite{carney_searches_2023,smolsky_direct_2024}. Recent work has demonstrated such kinematic reconstruction is possible for single nuclear $\alpha$ decays occurring in a micron-sized sphere, in which radon daughters have been implanted~\cite{wang_mechanical_2024}.
During the decay process, the emission of electrically charged decay products also induces discrete changes in the net electric charge of the microparticle through the emission both of the high-energy charged particles emitted by the decay, as well as any secondary emission of lower energy electrons or ions. These charge changes provide an additional signature of a single nuclear decay occurring in an electrically isolated particle, allowing individual decays to be identified with background levels $\lesssim$1~$\mu$Bq~\cite{wang_mechanical_2024}. A quantitative understanding of the net electric charge changes of such particles following nuclear decays, and their dependence on the type and energy of the emitted particles, is essential for understanding background levels that may be achievable in applications of levitated particles to neutrino physics. These techniques may also enable new radioassay methods for single dust-sized particles containing trace amounts of radioactivity, with potential applications in metrology~\cite{Fitzgerald_Primary_2025}, nuclear material characterization and analysis~\cite{Malyzhenkov_nuclear_2018}, and environmental monitoring of airborne radioactive particles~\cite{Gyoung_charging_2022,Kim_influence_2014}.

Previous studies have investigated secondary charge production by energetic particles incident on a variety of solid targets. For ions, electron yields, energy spectra, and angular distributions  have been studied for $\approx5~\mathrm{MeV}$ He\(^{2+}\) and \(\approx100~\mathrm{keV}\) heavy ions, which are the relevant energy scales for the $\alpha$ particles and nuclear recoils from radon progeny considered in this work (see, e.g.~\cite{koyama_secondary_1976, hasselkamp_particle_1992,Baragiola_principles_1993}). Secondary electron emission from incident electrons at the \(\approx1~\mathrm{MeV}\) energies relevant for the $\beta$ decays in the radon decay chain has also been studied~\cite{Schultz_secondary_1963}. For both ion and electron impacts, the total secondary electron yield is approximately proportional to the electronic stopping power of the incident particle, and depends on the transport and transmission through the surface~\cite{hasselkamp_particle_1992,Baragiola_principles_1993}.
Related measurements studying secondary electron emission from radioactive $\alpha$ sources embedded in metal surfaces have also been performed~\cite{anno_secondary_1963}, by measuring the total current generated from multiple decays. Most experiments have been performed on metallic targets, whereas dielectric materials present experimental challenges due to charge accumulation.

Radioactive decay-induced production of charged particles arising from the implantation of radon progeny into surfaces~\cite{Chernyak_Comprehensive_2023} has also been extensively studied in the context of background mitigation in rare event searches and neutrino physics. For example, in the KATRIN experiment implantation of radon daughters in detector surfaces leads to the production of secondary electrons and Rydberg atoms~\cite{Frankle_Radon_2011,Frankle_KATRIN_2022}, which are relevant for backgrounds in the neutrino-mass measurement~\cite{KATRIN_direct_2025}.
In gas and liquid phase time projection chambers (TPCs), low-energy backgrounds can be generated by decays of radon progeny embedded in detector surfaces, and characterizing low energy electrons and recoiling daughter nuclei produced by such backgrounds is important for searching for low-mass dark matter~\cite{aprile_light_2026,Aalbers_search_2023,Li_search_2023}. Implanted radon progeny can also be subsequently ejected from surfaces after further decays, and the techniques described here may be relevant for characterizing the escape of such daughter nuclei from surfaces at the single decay level.

In this work, we study the emission of primary and secondary charged particles in nuclear $\alpha$ and $\beta$ decays of $^{212}$Pb and its daughters shallowly implanted in a silica microparticle.
The measured net changes in the microparticle charge are correlated with the detection of the high-energy decay products incident on an external detector, composed of a scintillator read out by silicon photomultipliers (SiPMs). By measuring these coincident signals, the distributions of charge changes for $\alpha$ and $\beta$ decays are individually identified.

\section{\label{sec:methods}Methods}

\begin{figure*}[t!]
    \includegraphics[width=\textwidth]{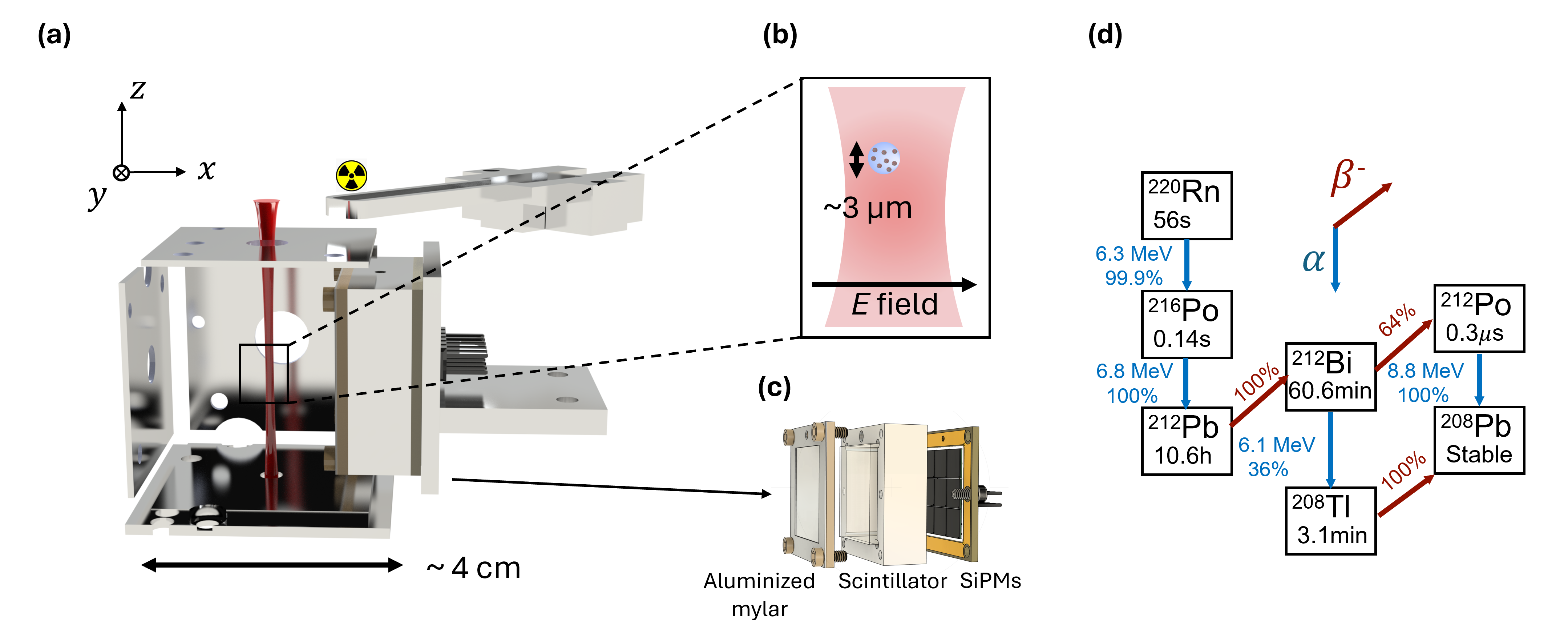}
    \caption{
    Overview of the experimental setup and radioactive decay chain.
    (a) Schematic view of the apparatus, showing the focused laser beam (red, not to scale), surrounding electrode structure, and the scintillator detector.
    A $2\,\mathrm{cm}\times2\,\mathrm{cm}$ scintillator, shielded with aluminized Mylar to suppress scattered light from the levitated particle, is positioned $\approx1\,\mathrm{cm}$ from the trapping region.
    The slide carrying radioactive spheres (whose location is denoted by the trefoil symbol) can be positioned above the beam to load spheres into the trap.
    (b) Schematic of a silica sphere carrying radioactive isotopes trapped by a focused laser beam with an electric field applied to monitor the net electric charge. 
    (c) Exploded view of the particle detector showing the entrance window, scintillator, and SiPM array.
    (d) $^{220}$Rn decay chain relevant to this work.
    Half-lives are indicated within each isotope box.
    Blue arrows denote $\alpha$ decays, with the dominant $\alpha$ energies and branching ratios indicated, while red arrows indicate relevant $\beta$ decays and their branching ratios.
    }
    \label{fig:scheme}
\end{figure*}

\subsection{\label{sec:ed}Experimental description}
Changes in the net electric charge of a microparticle associated with nuclear decays are measured with accuracy better than a single elementary charge using optically levitated silica spheres (diameter $\simeq 3~\mu\mathrm{m}$) held in high vacuum at pressures between $10^{-6}$ to $10^{-8}$~mbar. The spheres are implanted with $^{212}$Pb, which is an unstable daughter of $^{220}$Rn with a half-life of 10.6~hours~\cite{nudat}. Implantation is carried out in a dedicated vacuum chamber following the procedure described in~\cite{wang_mechanical_2024} by introducing $^{220}$Rn gas from an emanation source (Pylon TH-1025, 105 kBq in May 2024), which undergoes $\alpha$ decay to $^{216}$Po (see Fig.~\ref{fig:scheme}(d)), producing an ionized daughter a substantial fraction of the time when stopped in dilute gas~\cite{wang_mechanical_2024}. These ions are then drifted onto the surface of the sphere by a static electric field. Subsequent $\alpha$ decays of $^{216}$Po implant the recoiling $^{212}$Pb daughters in the surface of the silica sphere with a typical depth $\lesssim$60 nm~\cite{wang_mechanical_2024}.

Following implantation, the spheres are transferred into a high-vacuum chamber on a glass slide where they are loaded into an optical trap, consisting of a vertically oriented 1064~nm laser, which is weakly focused at the center of the vacuum chamber~\cite{wang_mechanical_2024}. Spheres are detached from the slide using a piezoelectric actuator to vibrate the slide while it is positioned above the trap focus. Many spheres are released from the surface and fall through the trapping region, from which a single sphere is captured. 

The slide, which usually has around $\approx$5-10 kBq activity directly after loading, is then retracted by 1~cm from its loading position above the optical trap using a vacuum translation stage, but remains within the vacuum system. When retracted, the intervening material between the slide and the scintillator detector (see Fig.~\ref{fig:scheme}(c)) consists of a 2~mm thick aluminum housing around the slide, a $1~\mathrm{mm}$-thick stainless-steel electrode, and a 4~mm thick aluminum housing surrounding the detector. This material is expected to fully absorb all $\alpha$ particles, while providing some degree of attenuation of $\beta$ particles. The low-energy $\beta$ spectrum of $^{212}$Pb is largely absorbed in the electrode or aluminum housings, whereas the higher-energy $\beta$ emissions from $^{212}$Bi and $^{208}$Tl can penetrate the material at certain angles. In addition bremsstrahlung can be generated from $\beta$ particles interacting with the intervening material. Therefore, residual radioactivity on the slide produces a background signal in the scintillator detector, as discussed in Sec.~\ref{sec:pd}. The measured background rate in the detector is $\approx$ 25 Bq approximately 5 hours after loading. In comparison, a typical trapped sphere exhibits an initial activity between $1-5$~mBq and yields 80--400 detected decay events before the activity has fully decayed away. Variations in the sphere activity can arise from the location of the sphere on the glass slide during loading, variations in charge buildup during loading, or other effects. However, such effects are not expected to affect the depth or distribution of the implanted $^{212}$Pb, and instead primarily affect the overall number of implanted atoms.

The optical trap is surrounded by five planar electrodes and a particle detector with a grounded housing, shown in Fig.~\ref{fig:scheme}(a). For charge measurements (see Sec.~\ref{sec:cc}), only the electrode facing the detector is biased, while all other electrodes, including the detector entrance surface, are held at ground. The electrodes incorporate apertures for the trapping and imaging beams, as well as ports for charge-control hardware, including an optical fiber coupled to a xenon flashlamp that produces ultra-violet (UV) light, and a heated filament. The sphere position is monitored using two orthogonal 532~nm imaging beams. The forward scattered light from the sphere is collected using lenses within the vacuum system focused on the trap location and imaged onto D-shaped mirrors that split the light onto balanced photodetectors (with one detector for each of the three axes). The differential signal on the balanced photodetector can be used to determine the sphere's displacement from the trap center in each direction. The displacement signal is sent to a Field-Programmable Gate Array (FPGA) based feedback system, which can output a real time signal to a piezo controlled deflection mirror that steers the trapping beam to provide feedback in the horizontal degrees of freedom, and an acousto-optic modulator (AOM) that modulates the trapping beam power to provide feedback to the vertical degree of freedom. In high vacuum, the center-of-mass motion of the sphere is stabilized using this active feedback~\cite{monteiro_optical_2017}. For data collection, the sphere's displacement signals and other experimental parameters are recorded by a data acquisition system (DAQ) while the triggered scintillator detector signal is recorded by a separate high-speed digitizer, as described in Sec.~\ref{sec:pd}.

\subsection{\label{sec:cc}Charge Detection}
\label{sec:charge_detection}
\begin{figure*}
    %units -e
    \includegraphics[width=\textwidth]{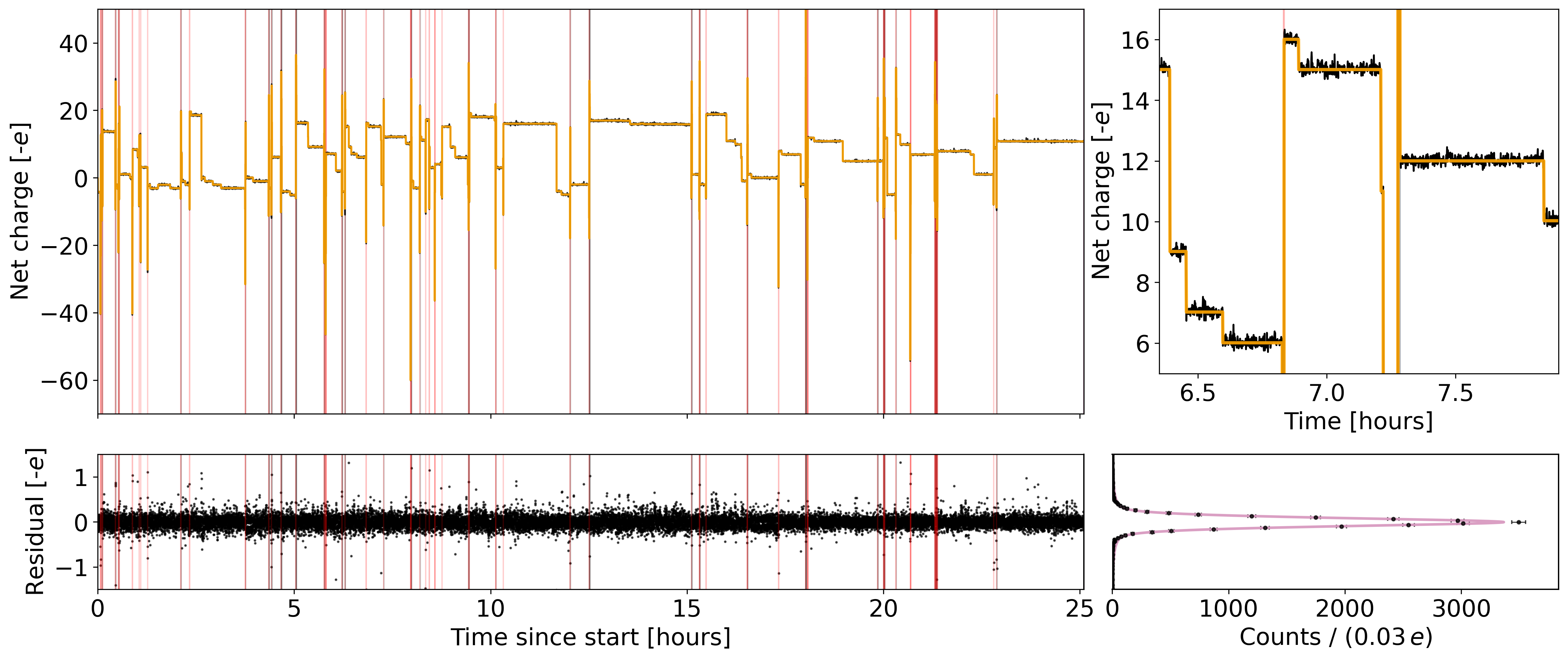}
    \caption{Charge detection for a radioactive sphere.
    Upper left: Measured net charge of a silica sphere implanted with $^{212}$Pb as a function of time.
    Net charge in units of $-e$ indicates the excess of electrons over protons in the sphere.
    The black curve shows the measured charge averaged in $\approx$3~s intervals, while the orange curve shows the best-fit charge trajectory (described in the main text) obtained after reconstructing the times of individual charge-change events.
    Light red vertical lines indicate periods during which electrons were added to the sphere using the filament, while gray shaded bands indicate ultraviolet illumination used to remove excess charge.
    Lower left: Residuals of the fit to the measured charge, demonstrating precise determination of the discrete charge state.
    Top right: A zoomed in view of the net charge of the silica sphere from 6.3 h to 8 h demonstrating the fit to several charge-change events.
    Bottom right: Histogram of residual with a Gaussian fit (pink).
    }
    \label{fig:lc}
\end{figure*}
After pumping to high vacuum, the net charge of the levitated sphere is determined from its driven response to an oscillating electric field. To neutralize the sphere, electrons can be added to the sphere via thermionic emission from a tungsten filament and removed using UV illumination~\cite{wang_mechanical_2024}. The response is calibrated in units of \(e\) using the observed discrete steps in the response amplitude from induced charge changes~\cite{wang_mechanical_2024}. The phase of the response relative to the driving electric field is used to determine the charge polarity. Each sphere is continuously monitored for \(1\) to \(3\) days after loading into the optical trap and reaching high vacuum, during which time the rate of charge-change events decreases following the half-life of $^{212}$Pb (10.6~h). The monitoring field is driven at a fixed frequency of \(197~\mathrm{Hz}\), which is well above the resonance frequency $\approx30~$Hz. The sphere’s charge is therefore measured by computing the Fourier transform of its displacement time series and extracting the amplitude corresponding to the monitor-field frequency and phase. Then, a piecewise-constant model \(f_i\) is fit to the measured data
\(q_i\) in each time bin \(i\). The step size and location of each charge change are determined by minimizing the total deviation between the fit and the data, together with a penalty term that favors the minimum number of charge changes, \(N\), required to describe the data. The fit is performed by varying the time and amplitude of each charge change to minimize the quantity $N+\sum_i |q_i-f_i|$.

A \(72~\mathrm{h}\) null test performed with non-radioactive spheres shows zero spontaneous charge-change events, placing an upper limit on the rate of backgrounds arising from random charging unrelated to radioactive decays in the particle of $<$12~$\mathrm{\mu Bq}$ at 95\% confidence~\cite{wang_mechanical_2024}, presenting a negligible background compared to the $\gtrsim$mBq rate of decays of implanted $^{212}$Pb and its daughters following the loading process described above. The observed charge changes for radioactive spheres are dominated by a net loss of electrons from the sphere, consistent with previous work~\cite{wang_mechanical_2024}. The scintillator detector is used to identify the high-energy charged particles emitted during these decay events when a time coincident signal is observed (see Sec.~\ref{sec:pd}).

This work reaches sphere activities up to an order of magnitude higher than those previously reported in Ref.~\cite{wang_mechanical_2024}, leading to a substantially improved measurement of the distribution of charge changes associated with decays within the particle. An example measurement of the observed charge changes over time for a single microsphere following loading with $^{212}$Pb is shown in Fig.~\ref{fig:lc}. To ensure the charge sensitivity is well below a single $e$, an electric field of \(7.5~\mathrm{kV/m}\) is applied to the sphere to monitor its net charge, $Q$, yielding a typical charge resolution of $\sigma_Q \approx 0.15\ e$ when the absolute value of the net particle charge is $<25\ e$. At this field strength, the charge resolution degrades when the net charge magnitude becomes larger, possibly due to fluctuations in the driving electric field, non-linearity in the detection, or other stray electric field noise. Therefore, an automatic discharge protocol is used to maintain \(|Q| < 25\,e\). The times at which the discharge protocol is active are recorded by the DAQ and are indicated in Fig.~\ref{fig:lc}. Dead time arising from the DAQ and discharging periods accounts for less than \(10\%\) of the total measurement time for all spheres considered.

Excluding periods during which the automatic discharge protocol is active, individual charge changes as small as \(1\,e\) are observed, while the largest events reach $(115 \pm 1)\,e$. A total of 66 events exhibit charge changes exceeding \(50\,e\). In addition to net electron loss, 33 events corresponding to a net loss of 1--2 positive charges are also observed.

\begin{figure}
    \includegraphics[width=\linewidth]{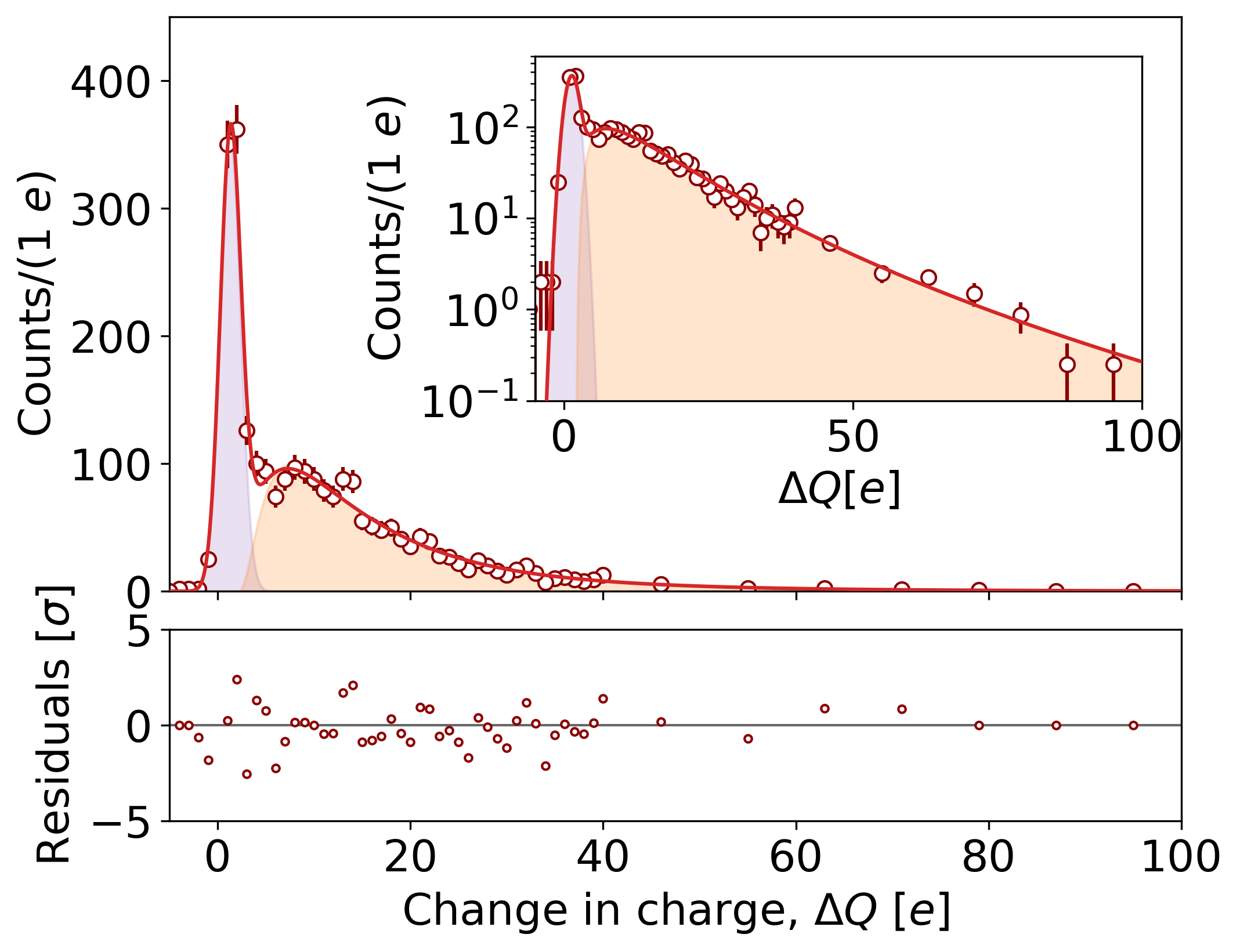}
    \caption{
    Distribution of reconstructed charge changes for all spheres (open markers). A fit (red line) to the sum of a Gaussian component (purple shaded) and a log-normal component (orange shaded) is also shown. The inset shows the same distribution on a log-scale. The bottom panel shows the residual between the data and fit, normalized by the Poisson counting error, $\sigma$.
    }
    \label{fig:cc}
\end{figure}

A histogram of all \(2602\) charge-change events recorded from the full set of 14 spheres is shown in Fig.~\ref{fig:cc}. 
Although the total activity of the measured spheres varies as described above, the distribution of charge changes was found to be statistically consistent for all spheres studied. 
We define the charge change from an initial net charge of the microsphere, $Q_i$, to final charge $Q_f$, $\Delta Q = -(Q_f - Q_i)$. Since nearly all observed decays produce a net loss of electrons, the sign is defined such that $\Delta Q$ is positive when more electrons are lost than protons. The distribution exhibits a sharp peak near \(\Delta Q = +(1.3 \pm 0.05)\,e\) that is modeled as a Gaussian distribution, as well as a pronounced high-charge tail that is empirically modeled by a log-normal distribution peaking at \(+(7.3\pm0.1)\,e\).

\subsection{\label{sec:pd}Particle detection}
A scintillator detector is used to identify $\alpha$, $\beta$, and $\gamma$ radiation emitted during decays within the particle, which can be measured in coincidence with the charge changes described in Sec.~\ref{sec:charge_detection}. This detector consists of a scintillator read out using a $3\times3$ array of SiPMs. This configuration allows both the energy deposited in the scintillator to be reconstructed (through the summed SiPM signal) as well as the position of the interaction to be determined (through the distribution of energy across the SiPM array). While future work will aim to reconstruct the momentum of the recoiling sphere in coincidence with a measurement of the momenta of the decay products using this position sensitivity~\cite{wang_mechanical_2024}, here we focus on differentiating $\alpha$ and $\beta$ decays only through the total energy deposited in the coincidence detector.

The scintillator material is a 20~mm\,$\times$\,20~mm$\,\times$\,5~mm gadolinium aluminum gallium garnet (GAGG) crystal optically coupled to nine 6~mm\,$\times$\,6~mm ONSemi MICROFJ-TSV-TR SiPMs mounted in a fully hermetic, polished aluminum housing. Each SiPM channel is read out individually by a Cremat CR-113 charge-sensitive preamplifier followed by a CR-200 shaping amplifier with a shaping time of $100~\mathrm{ns}$, integrated on a custom nine-channel board. The analog outputs are digitized using a CAEN DT5740 digitizer, which records 2~$\mu$s long waveforms for all channels when any SiPM's output voltage exceeds a predefined 100~mV trigger threshold (corresponding to more than $\approx 20$ photoelectrons detected on any channel).

Time coincidence between charge change and particle events is established using a hardware trigger issued by this CAEN digitizer routed to a dedicated input channel of the data-acquisition system (DAQ), which monitors the sphere’s displacement with a 0.1~ms time resolution. The data are synchronized to ensure the relative timing resolution between the data streams is equal to the coarser 0.1~ms time resolution of the DAQ, which is substantially faster than the $\gtrsim$~ms timescale at which charge changes can be resolved.

To suppress stray light scattered from the particle within the optical trap or other chamber surfaces and to allow the detector face to function as an electrode, two layers of aluminized Mylar (each 0.8 mg/cm$^2$ mass density per unit area) are mounted on the detector front surface. Two layers are used to minimize light leakage through pinholes in the aluminum coating while minimizing mass that can degrade the energy of incident $\alpha$ particles. When grounded together with the detector housing, this coating forms an effective shield for low frequency electric fields, preventing field leakage resulting from the SiPM bias voltages, which would otherwise induce significant forces on the microsphere as its charge changes throughout the measurement, resulting in drifts in the sphere’s equilibrium position. In all measurements reported here, the detector is positioned $9.6\pm0.1~\mathrm{mm}$ from the trapped sphere, subtending a solid angle of about $16\%$.

The scintillator detector is calibrated using a $^{212}$Pb source produced by implantation of activity directly onto a glass surface, following the same procedure used to implant activity to the microspheres themselves~\cite{wang_mechanical_2024}. During calibration, the source is placed within $0.1~\mathrm{mm}$ of the sphere’s trapping position relative to the detector, with the surface on which the activity was implanted directly facing the detector. Calibrations are taken under vacuum at a pressure of $\approx10^{-5}~\mathrm{mbar}$, where loss of energy due to gas collisions between the source and the detector is negligible. 

\begin{figure}
    \includegraphics[width=\linewidth]{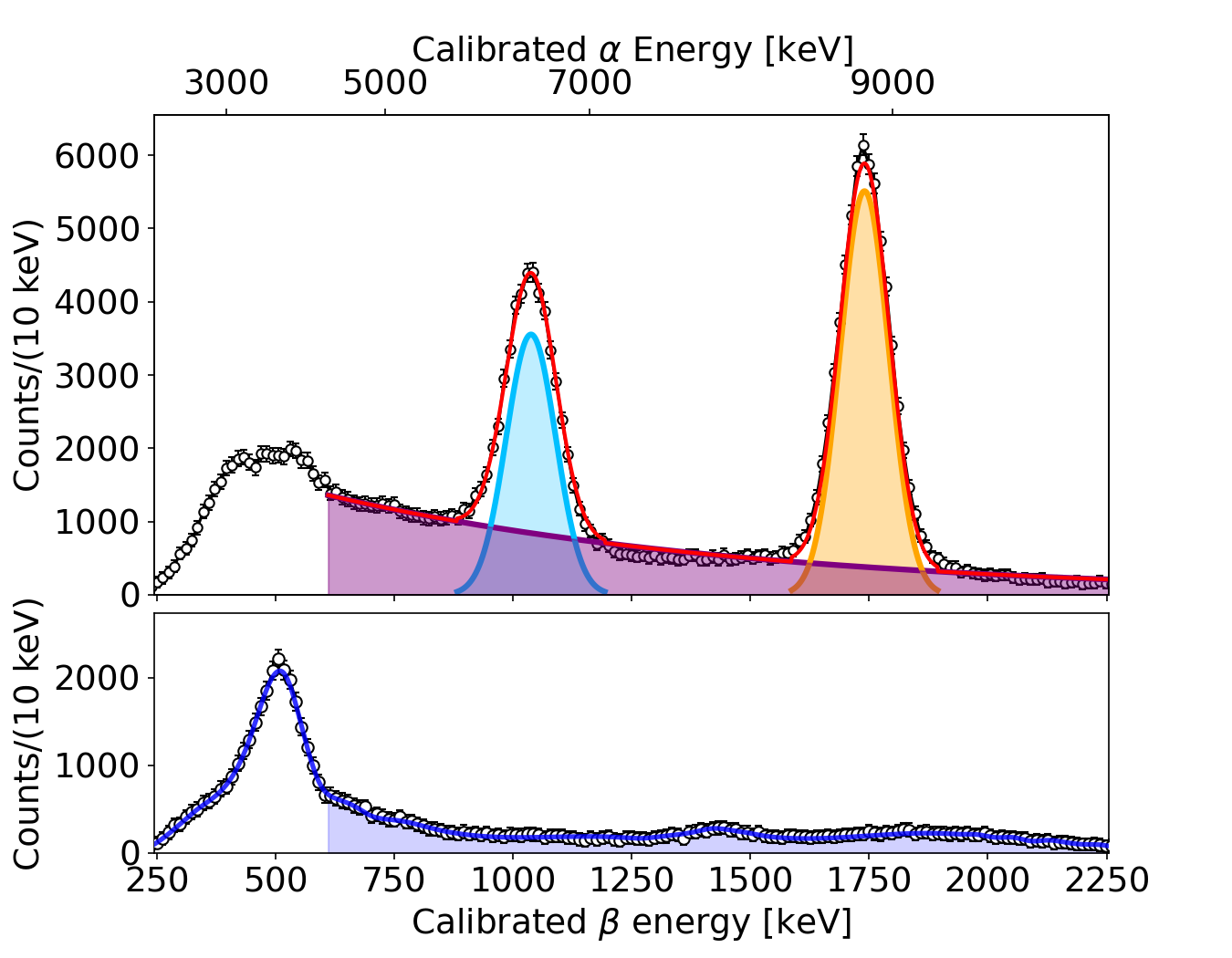}
    \caption{
    Calibration spectra measured with the scintillator detector.
    The top panel shows the spectrum acquired while a $^{212}$Pb source was positioned within $100\,\mu\mathrm{m}$ of the sphere trapping location in vacuum.
    The measured spectrum is shown by the open black markers, with the overall fit to the spectrum shown in red. The fit model includes $\alpha$ peaks from $^{212}$Bi (cyan) and $^{212}$Po (orange), as well as a slowly falling continuum above $\approx$500~keV from multiple $\beta$ decays, which is approximately modeled by an exponential dependence (purple, solid).
    The bottom panel shows the spectrum acquired in a $\approx$1~h exposure in the absence of a sphere, dominated by background events from the loading slide, which had an initial measured activity of $\approx4\times10^{3}\,\mathrm{Bq}$ for this dataset. The blue line shows the interpolation of this spectrum. The color-filled regions correspond to the spectral components used to model backgrounds in the coincidence analysis in Fig.~\ref{fig:sp}.
    }

    \label{fig:cal}
\end{figure}

Because the scintillation light yield is particle dependent and subject to quenching effects, interactions of $\beta$ and $\alpha$ particles yield different light responses for the same deposited energy, leading to a mis-reconstruction of the deposited energy if a calibration derived for a different particle species is applied~\cite{gironi_custom_2025,Furuno_2021}. In addition we observe a non-linear variation in the detector response for $\alpha$ particles at different energies, arising from energy dependence of this quenching and energy loss in the aluminized Mylar entrance window. After correcting for the observed energy dependence based on the measured $^{212}$Bi and $^{212}$Po peak locations to define the calibrated $\alpha$ energy scale, the energy resolutions are $\sigma_E = 230$$~\mathrm{keV}$ at the 6.1~MeV $^{212}$Bi peak and $\sigma_E = 133$~keV at the 8.8~MeV $^{212}$Po peak, as shown in the upper panel of Fig.~\ref{fig:cal}. The $\beta$ energy scale is calibrated by mono-energetic electron peaks (at 482~keV and 976~keV) from a $^{207}$Bi source. To reduce the dependence of the reconstructed energy on the detector position at which the particle interacts, the individual channel contributions are corrected by a calibration factor determined by minimizing the position-dependent energy spread of the $\alpha$ peaks from $^{212}$Bi and $^{212}$Po decays, which accounts for variations of $\approx40\%$ in the gain and light collection efficiency between channels.

The same reconstruction procedure used to define the calibrated $\beta$ and $\alpha$ energy scales from the calibration sources is then applied to data where a radioactive microsphere is present in the chamber. In addition to radioactive decays from the sphere, a significant background is present due to $\gamma$ and $\beta$ particles emitted by activity remaining on the dropper slide used for sphere loading, which remains inside the vacuum chamber during operation, as described in Sec.~\ref{sec:ed}. These particles can penetrate the trap electrodes and produce detector signals that are uncorrelated with charge-change events. The energy spectrum of this background is determined through measurements of the scintillator detector response in the absence of a trapped sphere, and is shown in the lower panel of Fig.~\ref{fig:cal}. The interpolated background shape is then used to model backgrounds arising from random coincidences of charge-change events with this background in Sec.~\ref{sec:tr}, based on a detailed timing analysis.

\section{Decay-resolved Coincidence Measurements}
\label{sec:coincidence}

Fourteen spheres in total were used to produce the charge-change spectrum in Sec.~\ref{sec:charge_detection}; however, only twelve had corresponding scintillator data and could be included in the coincidence analysis. We analyzed these twelve radioactive spheres trapped in the optical tweezer to study coincidences between discrete charge-change events and particle energy deposition measured by the SiPM array. 

\subsection{\label{sec:of}Time reconstruction of charge changes}
\begin{figure*}
    \includegraphics[width=\textwidth]{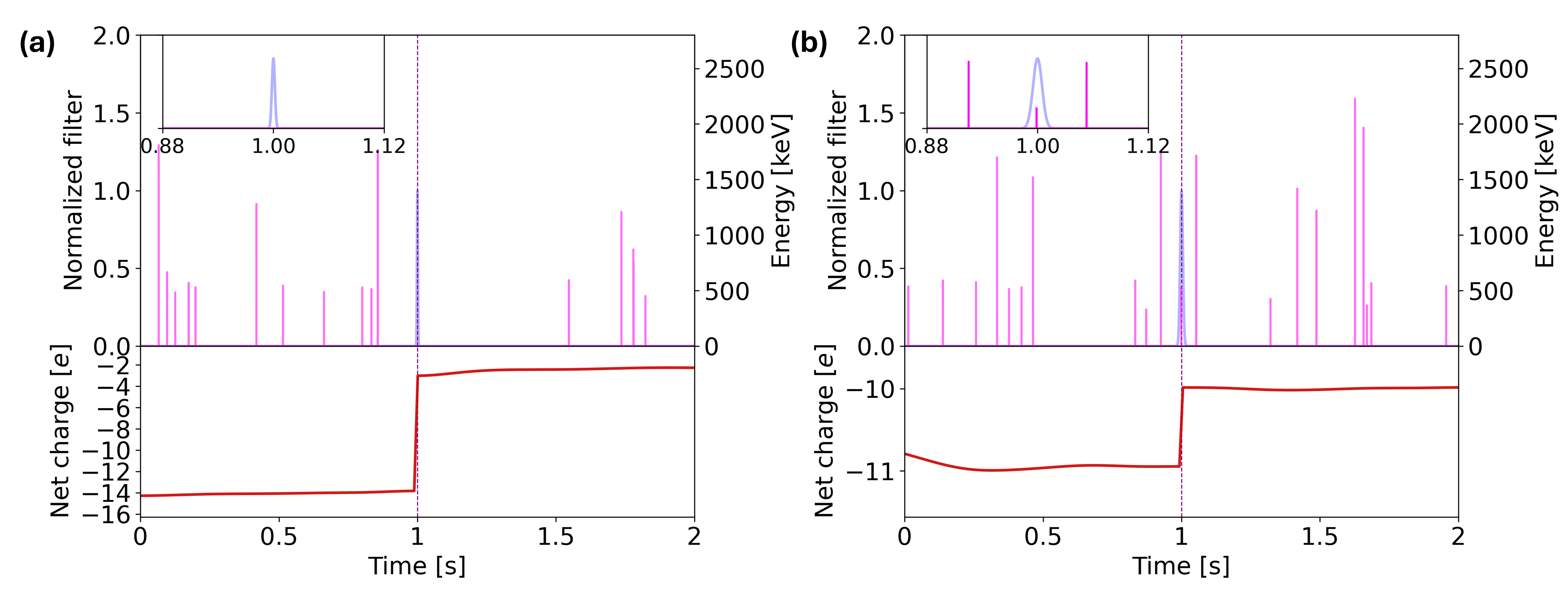}
    \caption{Comparison of charge-change event reconstruction and scintillator detector events.
    Two representative events with a non-coincidence (a) and a reconstructed coincidence between the charge change and scintillator (b) are shown. The upper panels show the scintillator reconstructed energy (pink) on the right axis, and the normalized optimal-filter response to the sphere charge change (blue) on the left axis, indicated by a Gaussian at the reconstructed charge-change time with the temporal resolution of the filter. The bottom panels show the net sphere charge versus time extracted at the drive frequency averaged over 25.6 ms (red). Insets show a magnified $\pm0.12~\mathrm{s}$ view around the reconstructed charge-change time.}
    \label{fig:op}
\end{figure*}
To reconstruct the times of charge changes within the DAQ data stream, a matched filter is implemented.
Matched filtering provides an optimal linear estimator for both the amplitude and time of the charge change based on the known signal shape and noise spectral density~\cite{wang_mechanical_2024, golwala_exclusion_2000,Schmerling_optimal_2025}. 
In our analysis, the signal template corresponds to the response of the oscillator to an abrupt change in electric force produced by a sudden change in the net charge of the sphere, evaluated at the applied electric driving frequency. When the sphere is driven by an electric field at a single frequency, the absolute value of the Fourier transform of the resulting displacement response to a step change in force can be described by a Lorentzian, with a small damping constant that accounts for phase fluctuations and the finite sampling frequency. The charge-change amplitude, $A_n$, at discrete time index $n$ can be estimated as:
\begin{equation}
A_n = \sum^{N-1}_{k=0} \frac{\tilde{s}^*_k \tilde{x}_k e^{2 \pi i k n/N}}{J_k}
\label{eq:optimal_filter_simp}
\end{equation}
where $k$ labels the discrete frequency index for a waveform with $N$ samples; $\tilde{x}_k$ is the Fourier transform of the measured sphere displacement; $\tilde{s}_k$ is the template of the sphere response near the driving frequency for a charge change of one elementary charge $e$, and $J_k$ is the noise power spectral density, which includes displacement noise at a level of $\approx10^{-11}$ m/$\sqrt{\mathrm{Hz}}$.

In practice, we apply this matched filter to the sphere position data within a narrow frequency band ($\approx$2~Hz) around the drive. The filtered waveform exhibits a sharply peaked response at the time of each charge-change event, arising from the change in the Coulomb force acting on the sphere. The location of the maximum of the filtered response for each reconstructed event identified in Sec.~\ref{sec:charge_detection} provides an estimate of the event time. 

The result of this timing reconstruction for two example events is shown in Fig.~\ref{fig:op}. Figure~\ref{fig:op}~(a) shows a charge-change event without a time coincident signal, consistent with a decay in the particle but where no signal is detected in the scintillator detector due to its finite solid angle. Figure~\ref{fig:op}~(b) demonstrates a temporal coincidence between a charge step measured by the sphere and an energy deposition event in the scintillator detector. The reconstructed charge-change time from the matched filter analysis, with resolution consistent with Fig.~\ref{fig:cr} is shown in the upper panels, while the bottom panels show the net charge estimated from the amplitude of the response at the drive frequency extracted from the Fourier transform in each 25.6 ms long window.

Finally, as shown in Fig.~\ref{fig:op}, because a relatively large electric field is applied to measure the charge changes, even a single-electron charge change produces a sufficiently strong impulsive response that the matched filter provides high temporal resolution.

\subsection{\label{sec:tr}Timing resolution}
\begin{figure}
    \includegraphics[width=\linewidth]{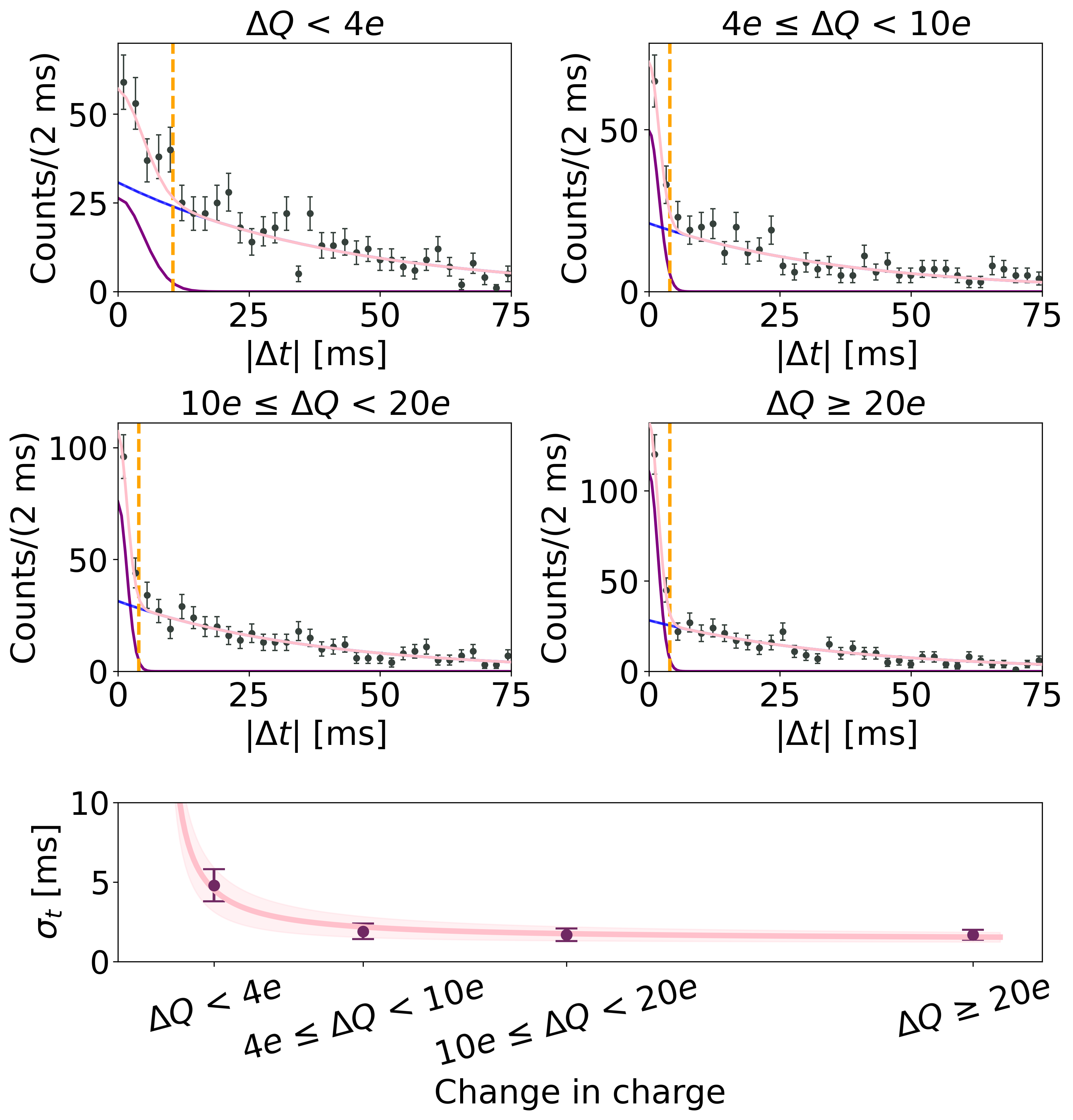}
    \caption{
    Time correlation between charge-change events and the scintillator detector.
    The four upper panels show the measured distribution of time differences, $|\Delta t|$ (black points), for events grouped by the magnitude of the observed charge change. This distribution is fit by a total model (pink) consisting of a Gaussian component centered at  $|\Delta t|$ =0 (purple) and an exponential background (blue).
    The vertical orange dashed line indicates the window used to identify  coincidences.
    The lower panel summarizes the extracted timing resolution as a function of charge-change magnitude, which is fit with a power-law dependence $\propto \Delta Q^{-1}$ (pink), where band denotes the $1\sigma$ uncertainty around the best fit.
    }
    \label{fig:cr}
\end{figure}
The time resolution of the measurement is limited by 
the resolution for reconstructing the charge changes from the sphere response, as described in the previous section. All other limitations on timing resolutions (due to, e.g., the DAQ bandwidth, scintillator detector response, escape of charged particles from the sphere, etc.) are estimated to be subdominant. 
Due to this finite time resolution, we associate events in which a charge-change signal is reconstructed at $t_m$ and a scintillator detector signal is measured at $t_0$ when they are coincident within:
\[
|t_0 - t_m| < t_{\mathrm{thr}}(\Delta Q),
\]
where $t_{\mathrm{thr}}$ is an empirically determined time window that depends on the amplitude of the charge change, $\Delta Q$. 

To determine $t_{\mathrm{thr}}$, the distribution of time differences, $\Delta t = |t_0 - t_m|$, between reconstructed charge-change events and the nearest scintillator detector trigger was analyzed. The resulting $\Delta t$ distribution consists of a narrow Gaussian peak superimposed on a smooth exponential background. The Gaussian component corresponds to true coincidences resulting from the same nuclear decay where both a charge change and a scintillator event are observed, while the exponential background arises from accidental coincidences of uncorrelated events. The accidental coincidence background is independently measured by generating a control sample with identical event statistics in which the charge-change times are randomized and the time differences to the nearest scintillator detector triggers are recomputed. This approach provides an empirical, high-statistics method of measuring the accidental background rate independent of true coincidences.

The background determined from this randomized timing data is plotted in Fig.~\ref{fig:cr} (blue lines), along with the observed event distribution. The events in the exponential accidental coincidence background arise from decays on the loading slide, and thus share the same energy spectrum as measured when no sphere is present in the bottom panel of Fig.~\ref{fig:cal}. The total event distribution is well described by this background (fully fixed in the model, with no free parameters) and a single Gaussian centered at $|\Delta t| = 0$, with the amplitude and resolution, $\sigma_t$, determined by a fit to the observed data.
The timing resolution, $\sigma_t$, varies from 4.9~ms for $\Delta Q<4~e$ to 1.6~ms for $\Delta Q \geq 20~e$, since larger charge changes can be reconstructed with better timing accuracy. 
To define coincidences in the following section, we take $t_\mathrm{thr} \approx 2\sigma_t$ in each bin of $\Delta Q$ used to analyze the energy spectrum of coincident events, i.e.
$t_{\mathrm{thr}}(\Delta Q < 4 e) = 10~\mathrm{ms}$ and $t_{\mathrm{thr}}(\Delta Q > 4e) = 4~\mathrm{ms}$. These thresholds 
give an expected efficiency to retain true coincidences of approximately 95\%.

The signal-to-background ratio for each bin of $\Delta Q$ can also be measured from the data in Fig.~\ref{fig:cr}. 
For $\Delta Q<4e$ the number of true coincidences is approximately 1.5$\times$ the number of accidental backgrounds, while for $\Delta Q>4e$ this ratio rises to $\gtrsim 3\times$.

\subsection{\label{sec:cs}Coincidence spectrum}
\begin{figure*}
    \includegraphics[width=0.9\textwidth]{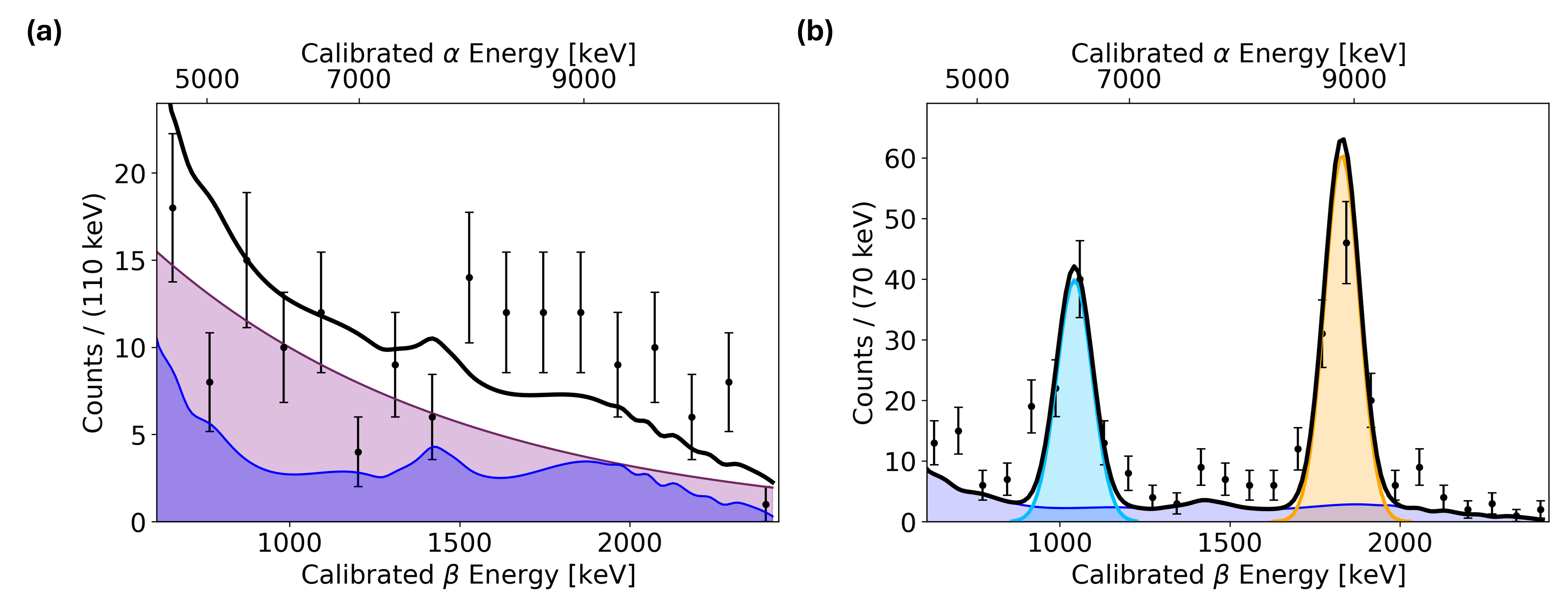}
    \caption{
    Energy spectra for reconstructed coincidences conditioned on charge-change amplitude.
    (a) Spectrum for $\Delta Q<4e$ (black points), compared to a total model (black line) consisting of the approximately exponential background from $\beta$ decays in the $^{212}$Pb calibration spectrum (purple) and the interpolated dropper background (blue).
    (b) Spectrum for $\Delta Q>4e$ (black points), where the total model (black line) now includes only the $\alpha$ components from the $^{212}$Pb calibration, as well as the interpolated dropper background (blue). The $^{212}$Bi and $^{212}$Po $\alpha$ peaks are indicated by the cyan and orange components, respectively. In both panels, each component indicated by the filled regions corresponds directly to the measured spectral shapes from the calibration in Fig.~\ref{fig:cal}.
    }

    \label{fig:sp}
\end{figure*}
Based on the timing analysis presented in Sec.~\ref{sec:tr}, the energy spectrum measured by the scintillator detector for events at a given point in the charge-change distribution can be measured. 
Events are divided into two categories to distinguish between the two features empirically observed in the charge-change distribution shown in Fig.~\ref{fig:cc}. The corresponding energy spectra for events satisfying these coincidence criteria are shown in Fig.~\ref{fig:sp}.

For small charge changes (\(\Delta Q < 4e\)), the energy spectrum in Fig.~\ref{fig:sp}(a) exhibits a broad continuum consistent with the sum of multiple $\beta$ and $\gamma$ decay branches in the \(^{212}\)Pb decay chain. As discussed in Sec.~\ref{sec:tr}, this spectrum is expected to arise from two sources: 1) true coincidences between a decay in the sphere and the detection of the emitted $\alpha$ or $\beta$ in the scintillator; and 2) accidental coincidences where a sphere decay results in a charge change, but the emitted $\alpha$ or $\beta$ is missed by the scintillator and an independent decay on the loading slide produces a scintillator signal within the coincidence window.

The energy distribution of the accidental coincidence background is modeled based on the measured spectrum for backgrounds originating from the loading slide when no sphere is present in the trap (see Fig.~\ref{fig:cal}~[bottom]). The normalization of this spectrum is fixed based on the integral of the exponential component measured in the timing analysis shown in Fig.~\ref{fig:cr} within the coincidence window (i.e., $<4$~ms for this charge-change bin). The above procedure allows the shape and normalization of the accidental coincidence background to be fully constrained with no other free parameters.

The shape and normalization of the true coincidences are also fully constrained using the $^{212}$Pb source calibration (Fig.~\ref{fig:cal} [top]). The spectral shape of the true coincidences is modeled by the exponential component describing the continuum of events around the $\alpha$ peaks in the $^{212}$Pb calibration data (purple solid line in Fig.~\ref{fig:cal} [top]). The normalization of this exponential component is then fixed by the integral of the Gaussian component from the timing analysis in Fig.~\ref{fig:cr} within the coincidence window. This procedure allows the observed spectrum to be adequately described with no free parameters, under the assumption that $\alpha$ decays do not significantly contribute to the $\Delta Q < 4e$ events.

For large charge changes (\(\Delta Q > 4e\)), the spectrum in Fig.~\ref{fig:sp}~(b) exhibits two distinct peaks whose energies are consistent with the calibrated \(\alpha\)-decay lines of \(^{212}\)Bi and \(^{212}\)Po. To model the observed spectrum, the accidental coincidence background is again fully fixed by the timing analysis, following the same procedure as for the $\Delta Q < 4e$ events. In addition to this background, a signal component is included based on the $^{212}$Bi and $^{212}$Po $\alpha$ components measured in the calibration spectrum shown in Fig.~\ref{fig:cal}~(top). The spectral shape (i.e., widths and relative amplitudes of the $\alpha$ peaks) and the total normalization are fixed by the calibration data and the timing analysis, respectively. However, the fit of the model to the data is improved by allowing a small scaling of the energy scale for the $\alpha$ peaks between the calibration data and charge-change data ($<5\%$), which is optimized as a single free fitting parameter. The best-fit systematic offset of approximately \(40~\mathrm{keV}\) is attributed to slow drifts in the detector energy calibration over the three-month data-taking period, while the calibration measurements were performed at the end of the experiment.

These results demonstrate that after accounting for accidental coincidence backgrounds, the $\Delta Q > 4e$ charge-change events are predominantly associated with \(\alpha\) decays, whereas $\Delta Q < 4e$ charge changes arise primarily from \(\beta\) decays. Unlike previous studies that reported only an average number of electrons removed per decay, our measurements reveal a broad distribution of charge ejection following individual \(\alpha\) decays. In particular, a single \(\alpha\) decay typically ejects approximately 10~electrons (and in rare cases more than 60 electrons) from the surface of the silica sphere. For detected energies within $2\sigma$ of the $^{212}\mathrm{Po}$ and $^{212}\mathrm{Bi}$ $\alpha$ lines, the charge-change distribution peaks at $\Delta Q = 10.2^{+2.2}_{-2.8}\,e$ and $9.2^{+2.3}_{-2.9}\,e$, respectively. Thus, in addition to the two protons escaping in the high energy $\alpha$ particle, $>$10 electrons are also typically ejected. In contrast, the $\beta$ decays that occur in the $^{212}$Pb decay chain typically eject $<4$ electrons, including the primary high-energy electron escaping from the decay.

Due to the timing resolution of the charge-change measurement, we do not separately distinguish the charge changes from the decay branch where $^{212}$Bi decays by $\beta$ emission followed by the $\alpha$ decay of its $^{212}$Po daughter, with a half-life of 0.3~$\mu$s. Based on the solid-angle coverage of our particle detector, events in which the $\beta$ emission is detected while the subsequent $\alpha$ decay of its $^{212}\mathrm{Po}$ daughter is missed are expected to contribute approximately $16\%$ of the observed coincidences in this portion of the decay chain. These events would provide a small additional contribution to the continuum in the $\Delta Q > 4e$ spectrum, but do not preclude the identification of large charge changes for $^{212}$Po where the $\alpha$ is detected by the scintillator.
\section{Summary and Discussion}
Using coincidences between a particle detector placed near a levitated silica microsphere with implanted radioactive isotopes and the change of the charge state of the sphere, we observe that different radioactive decay modes can be distinguished through the net charge change they induce. In particular, while both $\beta$ and $\alpha$ decays eject low-energy electrons in addition to the high-energy primary decay products, we directly verify that $\alpha$ decays occurring from radon daughters implanted $\lesssim$100~nm below the surface of the sphere produce substantially larger charge changes than $\beta$ decays. The distribution of charge changes associated with $\alpha$ emission is broad, with the largest observed event corresponding to a loss of 115 elementary charges. 

The substantially larger charge changes observed for $\alpha$ events can be understood as a consequence of near-surface secondary-electron emission driven by the large stopping power of the $\alpha$ particle and nuclear recoil in silica, which is much larger than that for MeV-scale energy $\beta$ particles exiting the microspheres~\cite{berger2005estar}. In particular, the shallow implantation depth can lead to events where the daughter nucleus recoils along a trajectory within tens of nanometers of the surface of the particle, permitting a large fraction of low-energy secondary electrons produced as the nuclear recoil stops to escape. The mean $\alpha$-correlated charge change observed here, of order ten elementary charges, is comparable to existing measurements of $\alpha$- and electron-induced secondary-electron emission from surfaces. For example, an average yield of $\approx 9.2$ electrons per $^{210}$Po $\alpha$ emerging from gold was reported in~\cite{anno_secondary_1963} and an average of $\approx 1.9$ electrons per $1\,\mathrm{MeV}$ electron incident on aluminum has been reported in~\cite{Suszcynsky_experiments_1991}. While in broad agreement with these average yields, these results resolve the details of the distribution of charge yields. Future work will study simulations of the low-energy processes needed to account for these observed distributions.

This work also demonstrates the ability to detect charged particles emitted in nuclear decays occurring in an optically trapped particle. In addition, while not used here, the position sensitivity of the scintillator detector allows both the energy and momentum of the decay products to be measured, allowing future work to study the kinematic reconstruction of the momentum of any undetected particles~\cite{carney_searches_2023}.

Beyond these applications, the ability to detect the emission of single, low-energy electrons with the techniques described here may provide a new approach for studying charge production by low energy nuclear recoils in solids, including rare low energy processes such as the Migdal effect~\cite{ibe_migdal_2018,aprile_search_2019,yi_direct_2026}.
Finally, in trapped-ion quantum processors, these results suggest that natural implantation of daughters of radon onto nearby surfaces or trap electrodes (during exposure of the components to room air during fabrication) can induce single nuclear decay events in which many low energy electrons may be simultaneously emitted and travel through the trapping region, interacting with the ions through Coulomb forces. In analogy to correlated errors observed in superconducting quantum processors arising from radioactivity, such events may become more evident as error correction techniques minimize the importance of error sources that are not correlated between qubits~\cite{Catelani_relaxation_2011,cardani_reducing_2021,vepsalainen_impact_2020}. Searches for millicharged particles or other rare-event searches using trapped ions may also eventually be sensitive to such low-energy charged particle backgrounds~\cite{Budker_millicharged_2022,Carney_trapped_2021,berlin_electric_2025}.

\begin{acknowledgments}
We thank the QuIPS team at Lawrence Berkeley National Lab for helpful discussions, and Gadi Afek and Fernando Monteiro for their early experimental efforts related to this work. This work was supported through the DOE Office of Science under Grants DE-SC0023672 and DE-SC0026367, and in part by ONR Grant N00014-23-1-2600. Y.-H. T. is supported by the Graduate Instrumentation Research Award (GIRA) from the Coordinating Panel for Advanced Detectors (CPAD).
\end{acknowledgments}

\bibliography{references_0}

\end{document}